\documentclass[prc,twocolumn,superscriptaddress,showpacs,preprintnumbers,amsmath,amssymb,nofootinbib]{revtex4-1}
\usepackage[dvips]{graphicx}
\usepackage{color}
\def\m@thcombine#1#2{%
  \setbox0=\hbox{$#1$}
  \setbox1=\hbox{$#2$}
  \ifdim\wd0>\wd1
    \setbox0=\hbox to\wd1{\hss\box0\hss}
  \else
    \setbox1=\hbox to\wd0{\hss\box1\hss}
  \fi
  \mathop{\vcenter{
    \offinterlineskip\box0\box1}}}
\def\lesim{\m@thcombine<\sim}
\def\gesim{\m@thcombine>\sim}

\def\vr{\mbox{\boldmath$r$}}

\def\+{\mbox{\unboldmath $+$}}
\def\-{\mbox{\unboldmath $-$}}
\def\={\mbox{\unboldmath $=$}}

\begin{document}
\title{Anderson-Bogoliubov phonon in inner crust of neutron stars: Dipole excitation in spherical Wigner-Seitz cell}
\author{Tsunenori Inakura}
\author{Masayuki Matsuo}
\affiliation{Department of Physics, Faculty of Science, Niigata University, Niigata 950-2181, Japan}

\begin{abstract}
\noindent {\bf Background:} The Anderson-Bogoliubov (AB) phonon, called also the superfluid phonon,
has attracted attentions since it may influence the thermal conductivity and other properties of inner crust 
of neutron stars.  However, there are limited number of microscopic studies of the AB phonon
where the presence of clusters is explicitly taken into account.\\
{\bf Purpose:} We intend to clarify how the presence of clusters affects the 
AB phonon in order to obtain microscopic information relevant to the coupling
between the AB phonon and the lattice phonon. \\
{\bf Methods:} The Hartree-Fock-Bogoliubov model
and the quasiparticle random-phase approximation 
formulated in a spherical Wigner-Seitz cell are adopted to describe neutron
superfluidity and associated collective excitations. We perform systematic numerical calculations
for dipole excitation
by varying the neutron chemical potential and the number of protons in a cell. \\
{\bf Results:} The model predicts systematic emergence of the  dipole AB phonon mode, which however 
exhibits strong suppression of phonon amplitude inside the cluster. We find also that the
phonon amplitude around the cluster surface varies as the neutron density.   
At higher neutron densities the AB phonon mode exhibits behaviour similar to the pygmy dipole 
resonance in neutron-rich nuclei.\\
{\bf Conclusions:} The dipole AB phonon mode does not penetrate into the clusters. 
This suggests that the coupling between the AB phonon and the lattice phonon may be
weak. 
\end{abstract}

\maketitle

\section{Introduction}

The inner crust of neutron stars is characteristic inhomogeneous
nuclear matter
that emerges in the surface region of a neutron star. In layers not very close to the
neutron star core, aggregates or clusters of neutrons and protons are formed in the environment 
of low-density neutron gas. 
The clusters, which form a Coulomb lattice,
resemble (but not identical to)  finite nuclei 
while the density of neutron gas varies from $\rho_n =0$
to $\sim 0.1 \rho_0$.  An important feature is that
neutrons in the inner crust are believed to become
superfluid at low temperatures due to the attractive nuclear force
in the $^1\mathrm{S}_0$ channel.

Some of the neutron star observables are linked to the static and
dynamic properties of the inner crust, characterized by the inhomogeneity 
and the superfluidity~\cite{Chamel-Haensel2008,Haensel-book,Pethick-Ravenhall95}. For example, 
the glitch is likely related to the pinning of superfluid
vortices~\cite{Anderson-Itoh1975,Alpar1977,Pines-Alpar92}. Cooling behaviours of young neutron 
stars~\cite{Lattimer94,Gnedin01} and soft X-ray transients~\cite{Shternin07,Brown09}
are governed by the heat capacity which is influenced strongly by the 
presence of the neutron pair gap~\cite{Pizzochero2002,Monrozeau07,Fortin2010}.
The quasiperiodic oscillations in giant flares are discussed in connection
with the lattice vibration (lattice phonon) of clusters~\cite{Duncan1998,Samuelson2007,Andersson09}. 
Recently another kind of dynamics, called the superfluid phonon
or the Anderson-Bogoliubov (AB) phonon, has attracted attention since it is  one of the
lowest frequency degrees of freedom other than the lattice phonon,
and it might influence thermal and mechanical properties of the inner 
crust~\cite{Aguilera09,Pethick10,Cirigliano11,Page-Reddy2012,Kobyakov13,Kobyakov14,Chamel13,Martin14}.

The superfluid phonon or the AB phonon is
a collective mode of excitation which emerges generally in 
neutral superfluid fermions~\cite{Anderson58,Bogoliubov59,Galitskii58}.  It is
 a Nambu-Goldstone mode associated with the gauge symmetry
broken spontaneously by the pair condensate. 
Its importance as heat carrier was first discussed by Aguilera \textit{et al.}~\cite{Aguilera09},
which points out possible new cooling mechanism effective
for neutron stars under the strong magnetic field and/or low temperature. An interesting feature is
that the AB phonon couples to the lattice phonon, and this coupling
is responsible to the description of the thermal conductivity~\cite{Aguilera09,Page-Reddy2012,Chamel13}
and  the quasi-periodic oscillations~\cite{Page-Reddy2012,Chamel13}.
It also might lead to formation of a new crystalline structure of the crust~\cite{Kobyakov14}. 
 
In the preceding works discussing the AB phonon in the inner crust
often adopted are macroscopic approaches based on effective field theory or superfluid
hydrodynamics~\cite{Aguilera09,Cirigliano11,Page-Reddy2012,Pethick10,Kobyakov13,Kobyakov14,Chamel13}.
In the macroscopic models, 
however, the AB phonon in uniform media is assumed, and 
the microscopic inputs are reflected only in model parameters. A complimentary approach 
would be many-body theoretical descriptions based on the nucleon degrees of freedom, and such
approaches may be useful to obtain
microscopic information on how the AB phonon mode couples to the lattice phonons,
and how the AB phonon mode evolves with the energy, the wave number, or the multipolarity.
Often adopted are
the selfconsistent nuclear density functional models such as the Hartree-Fock (HF)
or the Hartree-Fock-Bogoliubov (HFB) methods based on all nucleon degrees of freedom.
These models have been applied extensively to explore equilibrium crust 
configurations and to study pairing properties and the heat capacity of the 
matter~\cite{Barranco1998,Pizzochero2002,Sandulescu04a,Sandulescu04b,Baldo2005,Baldo2006,Monrozeau07,Fortin2010,Chamel10,Grill11,Pastore11,Pastore12}.
Collective excitation of inner crust matter has been studied also using the (quasiparticle) random phase
approximation (RPA, QRPA) formulated on the ground of the HF/HFB models~\cite{Khan05,Grasso08,Gori2004,Baroni2010,Martin14}. 
However, the AB phonon mode in the inner crust has been investigated very little
except  in Ref.~\cite{Khan05}, where  possible AB phonon mode is suggested for  a low-energy
 quadrupole excitation, called supergiant resonance. 
For uniform neutron matter, a microscopic study of 
the AB phonon mode is performed  in Ref.~\cite{Martin14},
using the density functional model and
the QRPA. 
The contribution of the AB phonon mode to the heat capacity is studied.
Comparison with the hydrodynamic description is discussed also.

In this paper we describe microscopically collective excitations of nuclear matter representing  
the inner crust of neutron stars. To describe the nucleon
many-body system and its excitations, we utilize a density functional model, i.e. 
the Skyrme-HFB method for the equilibrium configuration and the QRPA 
for small amplitude modes of excitation around the equilibrium~\cite{Nakatsukasa2016}. 
We neglect electrons for simplicity as they affects little neutron excitations such as the AB phonon mode. 
We adopt the Wigner-Seitz approximation
so that the calculation can be performed in a single spherical Wigner-Seitz cell.
In contrast to the previous works along the same line~\cite{Khan05,Grasso08,Gori2004,Baroni2010}, 
we pay special attention to the
neutron pair correlation and  we intend to reveal the properties of the AB phonon mode
realized in the inner crust.  As a first step of the study, we focus on the dipole excitation
in the present paper since this multipolarity is responsible for the coupling to the
the small amplitude displacement motion of the cluster, and hence to the
lattice phonon degrees of freedom.
In Sec. II, we briefly explain the model we use to describe the ground state and the excitation mode of the inner crust. In Sec. III, 
we discuss the static properties of the inner crusts, and the properties of the AB phonon mode with varying inner crust configurations.
Sec. IV is devoted to the conclusion.

\section{Models}

We employ the HFB theory and the QRPA in order to describe the static properties and excitation modes of the inner crust.
We neglect background electrons for simplicity and consider only zero temperature $T=0$.
The numerical code used in the present calculation is a revised version of the 
HFB+QRPA code developed in Refs.~\cite{Matsuo01,Matsuo05,Serizawa09,Matsuo10}
to describe isolated neutron-rich nuclei.  
To apply the model to the inner crust, we introduce the Wigner-Seitz approximation, i.e.
we treat a single cell of the lattice under suitable boundary conditions. A spherical 
 Wigner-Seitz cell is assumed, and we specify neutrons with chemical potential
$\lambda_n$ $(> 0)$ and protons with a fixed integer number $Z$ in the spherical box.
Following the standard prescriptions adopted in the HF/HFB 
calculations~\cite{Negele73,Barranco1998,Sandulescu04a,Sandulescu04b,Baldo2005,Baldo2006,Monrozeau07,Grill11,Pastore11,Pastore12}, we
 impose the Dirichlet-Neumann boundary 
condition~\cite{Negele73}; all even-parity wave functions vanish at the edge of the box, 
and first derivatives of odd-parity wave functions vanish at the edge of the box. 
Here we recapitulate the HFB+QRPA formalism briefly with emphasis on  
treatments relevant to the inner crust.  For other details of our HFB+QRPA approach, 
we refer the readers to Refs.~\cite{Matsuo01,Serizawa09,Matsuo10}.

\subsection{Hartree-Fock-Bogoliubov theory for static properties}

The HFB equation is solved in the coordinate representation. 
Employing the zero-range force, the HFB equation can be written as
\begin{eqnarray}
&& \sum_{\sigma^\prime}
\left(\begin{array}{cc}
h_\tau(\vr\sigma\sigma^\prime) - \lambda_\tau \delta_{\sigma\sigma^\prime} & \tilde{h}_\tau(\vr\sigma\sigma^\prime) \\
\tilde{h}^\ast_\tau(\vr\sigma\sigma^\prime) & -h^\ast_\tau(\vr\sigma\sigma^\prime) + \lambda_\tau\delta_{\sigma\sigma^\prime}
\end{array}\right) 
\phi_{i\tau}(\vr\sigma^\prime) \nonumber \\
&& \qquad= E \phi_{i\tau}(\vr\sigma)
\label{HFBeq}
\end{eqnarray}
where $\phi_{i\tau}(\vr\sigma)$ is $i$-th quasiparticle wave function with isospin $\tau=n,\, p$ 
and spin $\sigma= \uparrow,\,\downarrow$, and 
$E$ is quasiparticle energy. 
Using the spherical symmetry, we solve the HFB equation (\ref{HFBeq}) in the radial coordinate system. 
The HFB solution   is obtained for given proton number $Z$ and neutron chemical potential $\lambda_n$. 

The Hartree-Fock Hamiltonian $h_\tau$ (for the particle-hole channel) is derived selfconsistently
from  the zero-range Skyrme effective 
interaction. The adopted Skyrme parameter set SLy4~\cite{SLy4} is the one  which is 
adjusted to reproduce a theoretical equation of state of pure neutron matter~\cite{APR} and   
some fundamental experiential data of isolated nuclei in wide mass range, especially neutron-rich nuclei. 
The center-of-mass correction $m \to m \left( 1-\frac{1}{A}\right)^{-1}$ is not taken into account in the
present calculation. 
For the pairing potential $\tilde{h}_\tau$ (the pairing channel), we derive it from a 
density-dependent delta interaction (DDDI) of the form
\begin{eqnarray}
v_{\mathrm{pair},\tau}\left(\vr,\vr^\prime\right) = V_\mathrm{pair} \left[ 1 - \eta \left(\frac{\rho_\tau(\vr)}{\rho_c}\right)^\alpha \right]
\left( \frac{ 1 \- P_\sigma}{2}\right) \delta\left(\vr\-\vr^\prime\right) \nonumber\\
\label{DDDIpairing}
\end{eqnarray}
where $\rho_c=0.08$ fm$^{-3}$ and  $P_\sigma$ is the spin-exchange operator.
With this form, the pairing potential $\tilde{h}_\tau$ becomes a local pair potential
\begin{eqnarray}
\Delta_\tau(\vr) = V_\mathrm{pair} \left[ 1 - \eta \left(\frac{\rho_\tau(\vr)}{\rho_c}\right)^\alpha \right] \tilde{\rho}_\tau(\vr)
\label{pairing.Delta}
\end{eqnarray}
expressed with the pair density 
$\tilde{\rho}_\tau(\vr)=\langle \Psi_0 | \psi     (\vr\uparrow)   \psi     (\vr\downarrow) | \Psi_0 \rangle.$
The parameters are taken from Refs.~\cite{Matsuo07,Matsuo06}, where
the overall force constant $V_\mathrm{pair}= -458.4$ MeV fm$^{-3}$ is determined to reproduce the 
$^1\mathrm{S}_0$ scattering length $a=-18.5$ fm in the free space. 
The remaining parameters are adjusted as $\eta=0.845$ and $\alpha=0.59$ to reproduce the neutron mater pair gap obtained from the BCS calculation using a bare nuclear force~\cite{Matsuo06}.
Concerning the cutoff of the quasi-particle orbits, we set $j_\mathrm{max}= (75/2)\hbar$ for the single-particle partial waves and 
we introduce the cut-off quasiparticle energy $E_\mathrm{cut}=$ 60 MeV to avoid the ultraviolet divergence associated with the zero range pairing force.

\subsection{Quasiparticle random phase approximation}

The QRPA calculation  is performed in order to describe excitation modes 
built on top of the HFB ground state $\Psi_0$. The excitation modes are classified with
angular quantum numbers 
because of the spherical symmetry of the Wigner-Seitz cell. In the present study
we focus on the dipole excitations since the dipole multipolarity is
relevant to the coupling between the  
displacement motion of the cluster (or the lattice phonon) and the
other nuclear excitations. 

We utilize the linear response formalism~\cite{Matsuo01,Serizawa09,Matsuo10} for the QRPA. 
In order to explore the pairing collectivity, the AB phonon mode in particular, 
we describe  responses of the system with respect to not only 
the dipole moment operator $D$ but also the pair addition and removal 
operators $P_\mathrm{add}$ and  $P_\mathrm{rm}$, defined by 
\begin{eqnarray} 
&& D              = \sum_{\sigma} \int \!d\vr\, rY_{1M}(\hat{\vr})\, \psi^\dag(\vr\sigma)\psi(\vr\sigma)  \,, \nonumber \\
&& P_\mathrm{add} =         \int \!d\vr\,  Y_{1M}(\hat{\vr})\, \psi^\dag(\vr\downarrow)\psi^\dag(\vr\uparrow) \,, \nonumber \\
&& P_\mathrm{rm}  =         \int \!d\vr\,  Y_{1M}(\hat{\vr})\, \psi     (\vr\uparrow)  \psi     (\vr\downarrow) \,.
\label{eq:operators}
\end{eqnarray}
We solve the
QRPA linear response equations for fluctuations of the nucleon density $\rho(\vr)$, the 
nucleon pair density $\tilde{\rho}(\vr)$  and its complex conjugate $\tilde{\rho}^*(\vr)$. 
The spectral representation is adopted for the density response function and 
all the quasiparticle states used in the HFB calculation are included. 
We calculate the strength function 
\begin{eqnarray}
S(O;E) = \sum_{Mi} \delta(E - E_i) \left|\langle \Psi_i^{1M}| \hat{O} | \Psi_0\rangle\right|^2 \,,
\end{eqnarray}
for the operators $\hat{O}=D, P_\mathrm{add}$ and  $P_\mathrm{rm}$.
With  a small imaginary
constant $\epsilon$ in the energy argument, the 
delta function peaks in 
the strength functions are smeared with the Lorentzian function having
 the FWHM of $2\epsilon$. 
We evaluate the strength 
$B(O)=\sum_M\left|\langle \Psi_i^{1M}| \hat{O} | \Psi_0\rangle\right|^2$
of each excited state by integrating the strength function in an energy interval $E \in [E_i - 10 \epsilon, E_i + 10 \epsilon]$ around its peak energy $E_i$. We employ $\epsilon=10$ keV.
Three transition densities from the HFB ground state $\Psi_0$ to the $i$-th QRPA excited state $\Psi_i^{1M}$ 
\begin{eqnarray}
&& \delta       \rho _\mathrm{ph}(\vr) =   \langle \Psi_0 | \sum_\sigma\psi^\dag(\vr\sigma)     \psi     (\vr\sigma)     | \Psi_i^{1M} \rangle = Y_{1M}(\hat{\vr})\delta \rho _\mathrm{ph}(r)  \,,\nonumber \\
&& \delta\tilde{\rho}_\mathrm{pp}(\vr) = \langle \Psi_0 | \psi     (\vr\uparrow)   \psi     (\vr\downarrow) | 
\Psi_i^{1M}  \rangle = Y_{1M}(\hat{\vr})\delta\tilde{\rho}_\mathrm{pp}(r) \,,\nonumber \\
&& \delta\tilde{\rho}_\mathrm{hh}(\vr) =  \langle \Psi_0 | \psi^\dag(\vr\downarrow) \psi^\dag(\vr\uparrow)   | 
\Psi_i^{1M}  \rangle  = Y_{1M}(\hat{\vr})\delta\tilde{\rho}_\mathrm{hh}(r)\,,
\label{eq:densities}
\end{eqnarray}
are
obtained from the corresponding fluctuating densities at the peak energy $E_i$.
Note that all calculated spectra are discretized because of the boundary condition.

As the residual interaction  to be used in the QRPA calculation, 
we adopt the same effective pairing interaction, Eq.~(\ref{DDDIpairing}), in the particle-particle
and hole-hole channels. 
Concerning the residual interaction in the particle-hole channel, 
we adopt the Landau-Migdal approximation~\cite{Serizawa09,Matsuo10,Khan04,Khan09,Paar07} with a renormalization
scheme often employed in this approximation. Namely we replace the
self-consistent particle-hole interaction $v_{\mathrm{ph}}$ by the 
Landau-Migdal  interaction $f \times v_{\mathrm{LM}}$ derived from the Skyrme
interaction and renormalized with a factor $f$.
In describing isolated nuclei, this factor $f$ is fixed so that a
peak corresponding to displacement motion of the nucleus emerges
at zero energy. We adopt the same prescription in the present study.

\section{Results and discussion}

We have performed the HFB and QRPA calculations systematically for various configurations
obtained by changing the neutron chemical potential $\lambda_n$ and
the proton number $Z$ in order to discuss basic properties of the excitation modes of
nuclear matter in the inner crust,
without restricting ourselves to equilibrium configurations realized in 
realistic situation of
the inner crust.
The adopted proton numbers are $Z=20,28,40,50$,
chosen to cover the range  predicted in the previous HFB or HF calculations
for the equilibrium \cite{Negele73,Baldo2005,Baldo2006,Grill11}.
For the neutron chemical potential, we vary it in the range 
$\lambda_n = 1-6$ MeV, which corresponds to the density of neutron matter 
$\rho_n \approx 5 \times 10^{-4} - 1 \times 10^{-2}$ fm$^{-3}=( 3 \times 10^{-3}  - 7 \times 10^{-2} )\rho_0$.
 The box size, i.e. the radius of the Wigner-Seitz cell, is
fixed to $R_\mathrm{box}=20$ fm although
the box size $R_\mathrm{box}$ varies
if we find equilibrium configurations for different layers of crust.
An extension to a larger box size will be discussed in a forthcoming
paper. Since dependence on the proton number is weak as shown below, we mainly discuss
the case of $Z=28$ chosen as representative.

\subsection{Static properties}

\begin{figure}[tb]
\begin{center}
\includegraphics[width=0.499\textwidth,keepaspectratio]{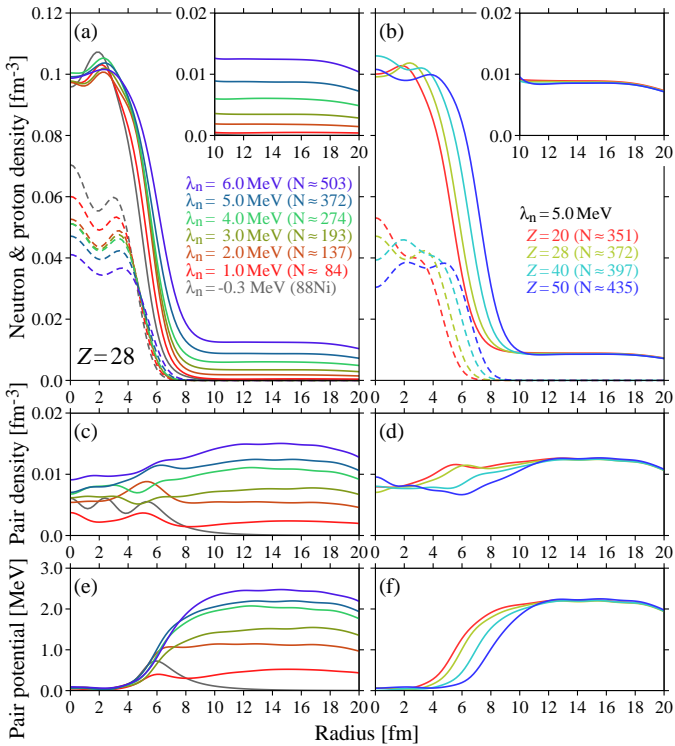}
\caption{(a) Calculated neutron densities (solid lines) and proton densities 
(dashed lines) for $Z=28$ system with $\lambda_n=$ 1.0 -- 6.0 MeV (red, brown, yellowish green, 
blue green, blue, and purple in order).
(b) The same but for $Z=$ 20 (red), 28 (light green), 40 (cyan), and 50 (blue) systems with fixed $\lambda_n=$ 5.0 MeV. 
(c)(d) Calculated neutron pair densities and (e)(f) neutron pair potentials for the same systems as (a)(b), respectively.
For comparison, densities and pair potentials of drip-line nucleus $^{88}$Ni are also plotted with gray line in (a)(c)(e).
}
\label{crust.density}
\end{center}
\end{figure}

Figure~\ref{crust.density}(a) shows the calculated neutron and proton densities, $\rho_n(r)$ and $\rho_p(r)$,
for $Z=28$ systems with $\lambda_n=$ 1.0 -- 6.0 MeV, and (b) for systems with different 
$Z$ with fixed $\lambda_n=$ 5.0 MeV.
For comparison plotted also are those for
$^{88}$Ni with $\lambda_n=-0.3$ MeV, the neutron drip-line isotope in the adopted HFB model.
From the neutron and proton densities one can see basic features of matter composition in the
inner crust: neutrons and protons aggregate together to form a cluster which resembles
a nucleus while the cluster is surrounded by low-density matter of neutrons.
The neutron density outside the cluster, i.e. $r \gesim 10$ fm
in Fig.~\ref{crust.density}(a), is almost independent on $r$. 
The neutron density outside the cluster increases with the neutron chemical potential
$\lambda_n$ as it should for the uniform neutron matter.
Note that a slight decrease of the neutron density
 around $r \sim R_\mathrm{box}$ is due to the specific choice of the
Dirichlet-Neumann boundary condition. 
If we take a different boundary condition where the two parity states are treated in the opposite way, 
the calculated density slightly increases around $r \sim R_\mathrm{box}$ in accordance with
Ref.~\cite{Pastore11}. In both cases, this artificial
effect of the boundary condition is small. 

In contrast to the simple trend of the neutron density outside the cluster,
density distributions in the region of the cluster behave in a different way.
Here we remark a few points. The increase of $\lambda_n$ does not increase the neutron central density of the cluster, but
it induces the increase of the radius of the cluster. 
Evaluating the surface radius of the cluster by fitting a function $\rho_\mathrm{cent}/\left( 1 + \exp[ (r - R) /a]\right) + \rho_\mathrm{matter}$ to the neutron and proton densities,
we have significant increase in the neutron surface radius
$R_n=$ 5.18, 5.57, 6.05 fm for $\lambda_n=$ 1.0, 3.0, 6.0 MeV, while
increase of the proton radius is smaller ($R_p =$ 4.83, 5.08, 5.48 fm in the same interval).
It is apparent that the neutron skin develops 
with increase of $\lambda_n$ as 
$R_n - R_p=$ 0.35, 0.49, 0.57 fm for $\lambda_n=$ 1.0, 3.0, 6.0 MeV.
Concerning dependence on the proton number $Z$ (see Fig.~\ref{crust.density}(b)), we find
increase of both the neutron and proton radii with increase of $Z$ while the
neutron density outside the cluster is determined solely by $\lambda_n$.
Note that the density profile of the cluster obtained in a similar 
Skyrme-Hartree-Fock calculation is analyzed in Ref.~\cite{Papakonstantinou13}
in terms of the equation of state of asymmetric nuclear matter. 

As a measure of the superfluidity, we show
the neutron pair density $\tilde{\rho}_n(r)$ in Figs.~\ref{crust.density}(c)(d),
and the neutron pair potential   $\Delta(r)$ 
in Figs.~\ref{crust.density}(e)(f).  The neutron pair density $\tilde{\rho}_n(r)$ 
exhibits a rather complex behaviour inside the surface of the cluster, but
it tends to converge to a constant value in external region $r \gesim 12$ fm far
beyond the cluster surface. 
It is noted that this
convergence is achieved by using a large cut-off value ($j_\mathrm{max}= (75/2)\hbar$ in the present calculation) 
of the angular momenta of the partial waves of the quasiparticle states. 

\begin{figure}[tb]
\begin{center}
\includegraphics[width=0.400\textwidth,keepaspectratio]{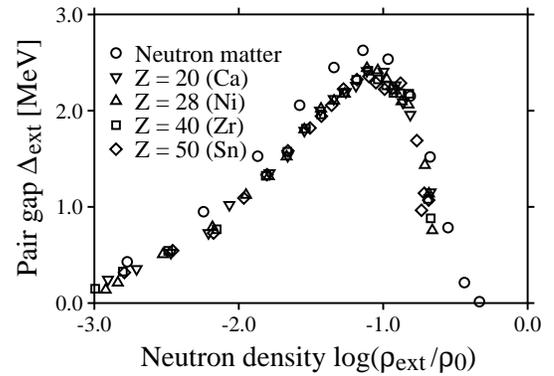}
\caption{Relation between neutron pair gap $\Delta_\mathrm{ext}$ 
and neutron density $\rho_\mathrm{ext}$
in the external neutron superfluid region
for $Z=$ 20 ($\bigtriangledown$),  28 ($\bigtriangleup$), 40 ($\square$), 
and 50 ($\diamond$) systems with $\lambda_n$ varied up to 10 MeV. 
Calculated pair gap of pure neutron matter is also plotted with circles for comparison.
}
\label{pairingGap}
\end{center}
\end{figure}

For quantitative assessment of the neutron pairing properties,  
we evaluate the average neutron density and the average pair gap 
in this converged region as
\begin{eqnarray}
\rho_\mathrm{ext}   = \frac{\int^{R_2}_{R_1} dr \, r^2 \rho_n(r)  }{\int^{R_2}_{R_1} dr \, r^2} \,,\quad
\Delta_\mathrm{ext} = \frac{\int^{R_2}_{R_1} dr \, r^2 \Delta_n(r)}{\int^{R_2}_{R_1} dr \, r^2}
\label{matter.Delta.rho}
\end{eqnarray}
with $R_1 =$ 12 fm and $R_2 =$ 18 fm.
Figure \ref{pairingGap} shows 
the relation between $\rho_\mathrm{ext}$ and $\Delta_\mathrm{ext}$, which are obtained  for
calcium ($Z=20$), nickel ($Z=28$), zirconium ($Z=40$), and tin ($Z=50$) systems with $\lambda_n$ 
varied  up to 10 MeV.
The obtained average gap $\Delta_\mathrm{ext}$ is insensitive to $Z$ and displays
a characteristic variation as a function of the neutron density  $\rho_\mathrm{ext}$: 
the maximum value $\Delta_\mathrm{ext} \sim$ 2.4 MeV at 
density $\log(\rho_\mathrm{ext}/\rho_0) \sim -1.0$ corresponding to $\lambda_n \sim 6$ MeV. 
For comparison, we show also the neutron pair gap of uniform neutron matter 
calculated with the HF+BCS methods using the same functionals (SLy4 and DDDI pairing of Eq.~(\ref{DDDIpairing})),
plotted with circles.
A good agreement with the neutron matter gap is seen, and it 
indicates that the pairing properties of neutrons
in the external region is  essentially the same as those of the
uniform neutron superfluid. 

The pairing properties inside and in the vicinity of the cluster
($r\lesim 12$ fm) is different
from those of the uniform neutron superfluid and from those of  isolated
nuclei as seen in Figs.~\ref{crust.density}(c)-(f).
Note that the Pippard's coherence length
$\xi=\hbar v_F/(\pi\Delta)$ in uniform neutron superfluid is estimated to be
order of 5 fm for $\rho_n/\rho_0 = 10^{-2}-10^{-1} $. 
The proximity effect \cite{Pizzochero2002, Barranco1998} is 
expected for the region $r\lesim 12$ fm.

\subsection{Collective dipole modes of excitations}

\begin{figure}[!tb]
\begin{center}
\includegraphics[width=0.499\textwidth,keepaspectratio]{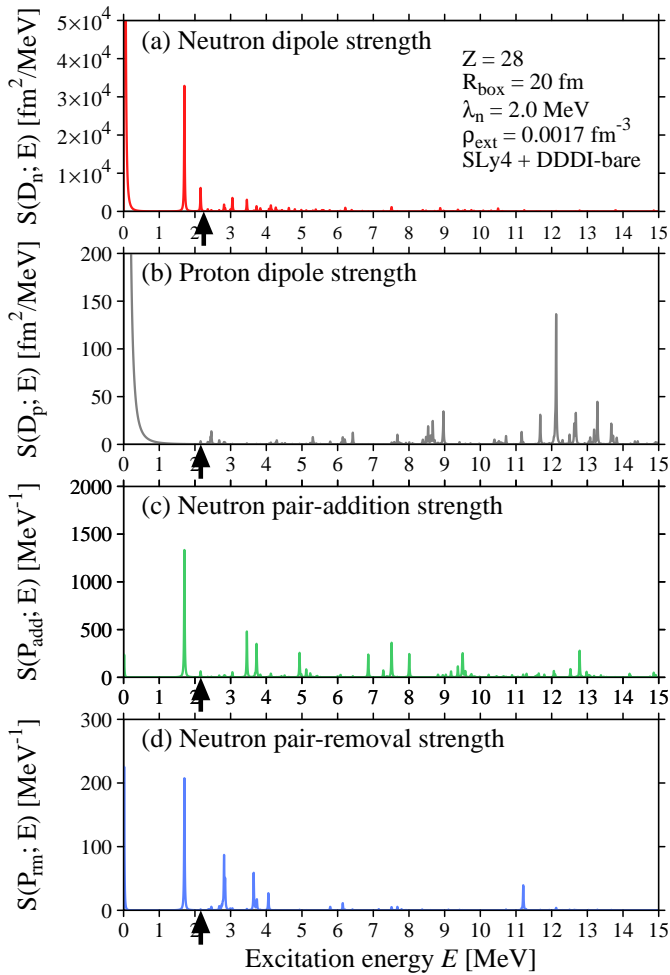}
\caption{(Color online) (a) Strength function $S(D_n;E)$ for neutron dipole operator, 
(b) $S(D_p;E)$ for proton dipole operator, (c) $S(P_\mathrm{add};E)$ for neutron pair-addition operator  
and (d) $S(P_\mathrm{rm};E)$ for neutron pair-removal operator, calculated 
for $Z=28$ system with $\lambda_n =$ 2.0 MeV,
plotted as a function of the excitation energy $E$.
Arrows indicate the threshold energy $2\Delta_\mathrm{ext}$. 
The smearing parameter is $\epsilon=$ 10 keV. 
}
\label{neutron.proton.Padd.Premove}
\end{center}
\end{figure}

We shall now discuss excitation modes in the inner crust, especially those of
the dipole character.
Figures~\ref{neutron.proton.Padd.Premove}(a)-(d) show the strength functions 
$S(O;E)$ for the neutron dipole
moment, $O=D_n$, the proton dipole moment  $D_p$, 
the neutron pair-addition operator $P_\mathrm{add}$ and the neutron pair-removal operator $P_\mathrm{rm}$ 
for the $Z=28$ system with 
$\lambda_n = 2.0$ MeV. 

\begin{figure}[!tb]
\begin{center}
\includegraphics[width=0.499\textwidth,keepaspectratio]{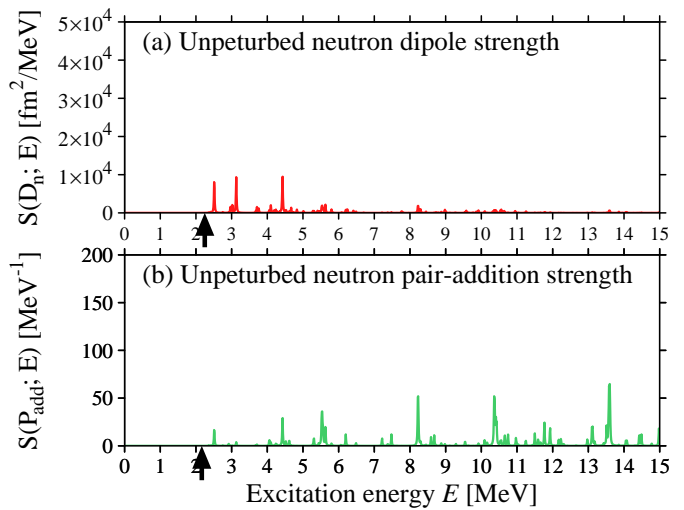}
\caption{(Color online) (a) Unperturbed strength function $S(D_n;E)$ for neutron dipole operator and (b) $S(P_\mathrm{add};E)$ for neutron pair-addition operator, 
calculated for $Z=28$ system with $\lambda_n =$ 2.0 MeV.
The scale of vertical axis of $S(P_\mathrm{add};E)$ is ten times larger than that of Fig.~\ref{neutron.proton.Padd.Premove}(c).
}
\label{unperturbed}
\end{center}
\end{figure}

For comparison, we show in Fig.~\ref{unperturbed} the unperturbed strength functions for the neutron
dipole operator $D_n$ and the neutron pair-addition operator $P_\mathrm{add}$,
i.e. those calculated for unperturbed neutron two-quasiparticle configurations.
It is seen that the low-energy peaks at $E=1.71$ MeV and $E=0$ in Fig.~\ref{neutron.proton.Padd.Premove}
 have strengths much larger
than those of the unperturbed two-quasiparticle configurations. We note also 
that these peaks are located below twice of the neutron pair gap $2\Delta_\mathrm{ext}$
(indicated with arrows), and are well separated from densely distributed states, which
would become continuum states if we take the limit of infinite $R_\mathrm{box}$.
These observations suggest collective nature of these low-lying peaks.

\begin{figure}[!tb]
\begin{center}
\includegraphics[width=0.470\textwidth,keepaspectratio]{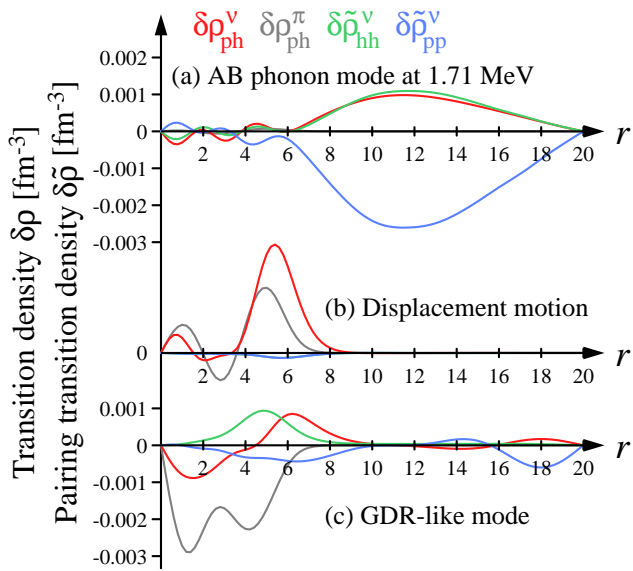}
\caption{(Color online) (a) Transition densities  of  the AB phonon mode at $E=1.71$ MeV  obtained for $Z=$ 28 
system with $\lambda_n=$ 2.0 MeV. Red, gray, green, and blue lines correspond to $\delta\rho^\nu_\mathrm{ph}$, $\delta\rho^\pi_\mathrm{ph}$, 
$\delta\tilde{\rho}^\nu_\mathrm{pp}$, and $\delta\tilde{\rho}^\nu_\mathrm{hh}$, respectively. 
(b) The same but for the displacement motion of cluster at $E=0$. The scale is in
arbitrary unit. (c) The same but for the GDR-like mode, calculated 
at $E=12.2$ MeV with employing a large smearing width $\epsilon=500$ keV.
}
\label{CM.ABmode.LED}
\end{center}
\end{figure}

Let us examine  the peak at $E=$ 1.71 MeV.
It has large strengths for  the neutron dipole moment $D_n$,
the neutron pair-addition operator $P_\mathrm{add}$ and the neutron pair-removal operator 
$P_\mathrm{rm}$. This state is the largest, apart from the peak at $E=0$ (discussed below), and exhausts 
 40 \%, 11 \%, and 25 \%, respectively, of the total strengths (non-energy-weighted sums integrated 
up to 50 MeV excluding the zero-energy state). 
The neutron dipole strength $B(D_n)$ of this state  is
enhanced due to the QRPA correlation by a factor of about ten
in comparison with strengths of the unperturbed two-quasi-neutron configurations.
 More prominent is the
pair-addition and removal strengths, $B(P_{\mathrm{add}})$ and $B(P_{\mathrm{rm}})$, in which 
drastic enhancement of order of $10^2$ relative to the unperturbed strengths is seen. 
We also note that
 these strengths are much larger than those of isolated 
nucleus by a factor of $10^{2-3}$. In contrast, the proton dipole strength of the
1.71 MeV state is negligibly small.
These suggest that the 1.71 MeV state is related to excitation of neutron superfluid 
rather than excitation of the  cluster. In fact, this character is clearly seen in
the transition densities of this state, shown
in Fig.~\ref{CM.ABmode.LED}(a). We will discuss detailed properties of this mode later. 

The peak at $E=0$ is extremely intense in the neutron and  proton dipole strength
functions $S(D_n;E)$ and $S(D_p;E)$. The transition densities of this mode is shown in 
Fig.~\ref{CM.ABmode.LED}(b): Both the neutron and the proton particle-hole transition densities
  $\delta\rho^\nu_\mathrm{ph}(r)$, $\delta\rho^\pi_\mathrm{ph}(r)$ have large
amplitudes in the surface area of the cluster, $r \approx$ 4 -- 8 fm, and their
shapes resemble to the derivatives $d\rho_n(r)/dr$
 and $d\rho_p(r)/dr$ of the densities.
All these features indicate 
displacement motion of the cluster as the character of this mode.
The displacement motion is expected to emerge at zero energy if the HFB+QRPA
calculation is fully selfconsistent and the box size is infinite. In the present calculation, 
we force it to be located at zero by adjusting the renormalization factor ($f=$ 0.825 in this case). 
A bunch of large proton strength in $S(D_p;E)$  around 12 MeV is excitation similar to the giant dipole resonance 
(GDR) in isolated nuclei. However, this GDR-like mode exhibits some differences from
the GDR.
For example, the neutron transition densities
have finite amplitudes in the external region, shown in 
Fig.~\ref{CM.ABmode.LED}(c), which indicates
 that the GDR-like mode
is coupled to excitations of surrounding neutron superfluid.

\subsection{Anderson-Bogoliubov phonon mode}

Characteristic features of the 1.71 MeV state are seen in the transition densities  
(Fig.~\ref{CM.ABmode.LED}(a)). The most noticeable is common sinusoidal 
oscillatory behaviour with long wave length seen in the
the three neutron transition densities 
$\delta\rho^\nu_\mathrm{ph}$, 
$\delta\tilde{\rho}^\nu_\mathrm{pp}$, and $\delta\tilde{\rho}^\nu_\mathrm{hh}$
in the region $r \gesim 6$ fm  outside the cluster.
It is  noticed also that the pair-addition and -removal transition densities,
$\delta\tilde{\rho}^\nu_\mathrm{pp}$ and $\delta\tilde{\rho}^\nu_\mathrm{hh}$,
are out of phase. 
It is recalled here that the AB phonon mode is
essentially the small-amplitude oscillation of the pair gap and the pair condensate, $\Delta(\vr,t)\sim
\Delta e^{i\varphi({\bf r},t)}, \tilde{\rho}(\vr,t)\sim \tilde{\rho} e^{i\varphi({\bf r},t)}$ with
the phase oscillation $\varphi(\vr,t)=\epsilon e^{i {\bf q}\cdot{\bf r} -i\omega t}$. The collective motion
 induces fluctuating pair densities $\delta\tilde{\rho}(\vr,\omega) \sim i\epsilon\tilde{\rho} e^{i {\bf q}\cdot{\bf r}}$
 and $\delta\tilde{\rho}(\vr,\omega)^*\sim-\delta\tilde{\rho}(\vr,\omega)$. 
Also fluctuation of the particle density is induced simultaneously
$\delta\rho \propto \delta\tilde{\rho}\times (\hbar\omega/e_F)(\rho/\tilde{\rho})$
due to the hydrodynamic character of the AB phonon mode~\cite{Martin14,Pethick-Smith2002}. 
These characteristic signatures of the AB phonon mode are observed in  the  
 three transition densities 
$\delta\rho^\nu_\mathrm{ph}$, 
$\delta\tilde{\rho}^\nu_\mathrm{pp}$, and $\delta\tilde{\rho}^\nu_\mathrm{hh}$ of
the 1.71 MeV state, especially in the region of neutron superfluid, and hence we
interpret it as the AB phonon mode realized in the inner crust.

It should be noticed also that the characters as the AB phonon mode is seen
only in the region of neutron superfluid $r \gesim 6$ fm outside the cluster, and
behaviour of the transition densities in the cluster region $r\lesim 6$ fm is very different from
that in the external region.
All the neutron transition densities here are much smaller than those outside, and wiggle with short wave lengths.
If it were the AB phonon realized in uniform superfluid, the lowest-energy (lowest harmonics) dipole eigen mode 
under the present Wigner-Seitz approximation would be  a standing wave having a radial
profile  $\propto j_1(qr)$ (the spherical Bessel function) with nodes at $r=0, \, R_{\mathrm{box}}$.
The AB phonon mode in the present case display approximate node at 
 $r\approx 6.5$ fm located in the skin region
of the cluster, and essentially vanishing amplitude inside the cluster. This indicates that
 the AB phonon mode
in the present calculation does not penetrate into the cluster.  The dipole AB mode 
appears as if it is the standing wave taking place in the region 6.5 fm $\lesim r < R_{\mathrm{box}}$.

\begin{figure}[!tb]
\begin{center}
\includegraphics[width=0.470\textwidth,keepaspectratio]{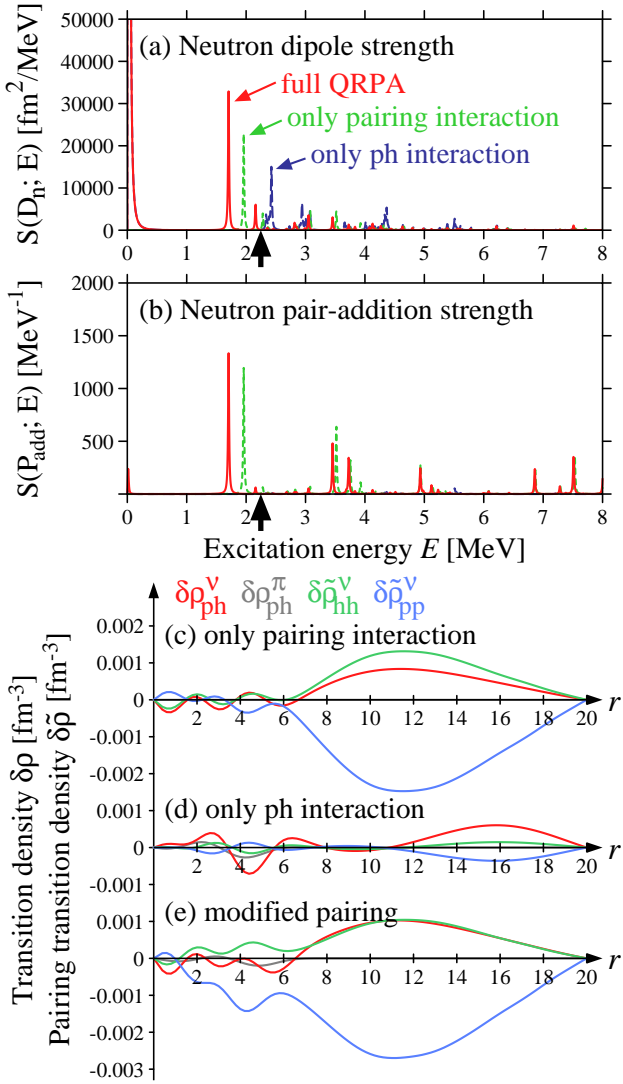}
\caption{(Color online) 
Uppers: (a) Strength function $S(D_n;E)$ and (b)  $S(P_\mathrm{add};E)$ 
 calculated with full QRPA residual interaction (red), only residual pairing interaction (green dashed), 
and only residual particle-hole interpretation (blue long-dashed). 
The system is for $Z=28$ and $\lambda_n = 2.0$ MeV.
Lowers: Transition densities of the largest peak in calculations with 
(c) only residual pairing interaction, (d) only residual particle-hole interaction, 
and (e) modified pairing interaction.
Red, gray, green, and blue lines correspond to $\delta\rho^\nu_\mathrm{ph}$, $\delta\rho^\pi_\mathrm{ph}$, 
$\delta\tilde{\rho}^\nu_\mathrm{pp}$, and $\delta\tilde{\rho}^\nu_\mathrm{hh}$, respectively. 
See text for details.
}
\label{residual}
\end{center}
\end{figure}

In order to investigate the origin of the characteristic features of this excitation mode, we 
look into roles of the residual interaction. For this purpose we shall drop off or modify some parts of the
residual interaction which enters in the QRPA calculation. 
Figures~\ref{residual}(a)(b) show 1) the strength functions $S(D_n;E)$ and $S(P_\mathrm{add};E)$ obtained 
with only the residual pairing interaction (i.e. the particle-hole part of the residual interaction is neglected)
plotted with green curves, and 2) those (blue curves) obtained with only the residual particle-hole interaction 
(the residual pairing interaction is neglected instead). They are compared with the full QRPA result (red curves), 
the same as shown in Fig.~\ref{neutron.proton.Padd.Premove}(a).  It is seen that the calculation using only the residual pairing interaction gives the
result which is similar to the full QRPA calculation with small changes 
in the excitation energy (increase to $E=1.96$ MeV) and in the strengths (reduction of 10 -- 30 \%). 
In fact, the transition densities in this calculation, shown in
Fig.~\ref{residual}(c), also resemble those of the full calculation 
(Fig.~\ref{CM.ABmode.LED}(a)) with small reduction of the
neutron particle-hole transition density. 
If the residual pairing interaction is neglected (and only the particle-hole
interaction included), however, the result is completely different. 
The excitation energy is moved upward further, 2.45 MeV, and
the  strength  $S(P_\mathrm{add};E)$ becomes negligibly small. The feature of the long-wave-length
oscillation of the neutron superfluid is completely lost as seen in the transition densities shown in 
Fig.~\ref{residual}(d). We thus see that the characteristic feature of the 1.71 MeV state is caused by the
pairing correlation, and we confirm that it is essentially the AB phonon mode of the neutron superfluid

We performed also another modified QRPA calculation where the residual pairing interaction is 
changed to a simple density-independent contact interaction,
which is produced by replacing the density $\rho_n(\vr)$ in the density-dependent factor 
$ \left[ 1 - \eta \left(\frac{\rho_n(\vr)}{\rho_c}\right)^\alpha \right]$ 
in Eq.~(\ref{DDDIpairing})  with the neutron superfluid density 
$\rho_\mathrm{ext}$. This has stronger
interaction strength inside the cluster than the original one while the interaction strength in 
the neutron superfluid region is unchanged.
In this calculation the lowest excited state (excluding the displacement motion) appears 
at $E=1.62$ MeV with small difference from the original value $1.71$ MeV. Influence of
the modified pairing interaction is seen in the transition densities in
the region $r \lesim 6$ fm, see Fig.~\ref{residual}(e): The neutron pair transition densities 
$\delta\tilde{\rho}^\nu_\mathrm{pp}(r)$ and $\delta\tilde{\rho}^\nu_\mathrm{hh}(r)$
are enhanced inside the cluster as is expected from the increased pairing
interaction in this region. 
It is noted however that the pair transition densities  
$\delta\tilde{\rho}^\nu_\mathrm{pp}(r)$ and $\delta\tilde{\rho}^\nu_\mathrm{hh}(r)$
wiggle there (suggesting single-particle character), and the amplitude of
the neutron particle-hole transition
density $\delta\rho^\nu_\mathrm{ph}(r)$ remains small. 
The hydrodynamic AB phonon mode is not realized inside the cluster even with
this modified pairing interaction.

We here point out that if we consider the
coherence length $\xi=\hbar v_F/(\pi\Delta)$ of neutron pair correlation
using the Fermi velocity $v_F$
and the pairing gap $\Delta$ appropriate for inside the cluster
($\hbar v_F \approx 60$ MeVfm, $\Delta\approx 0.2$ MeV), it is estimated to
be $\xi_\mathrm{int} \approx 100$ fm, much larger than the size $R_\mathrm{cluster}$ 
of the cluster. In this 
situation the hydrodynamic feature of the AB phonon mode is hardly realized as
finite-size effects might dominate there because of $R_\mathrm{cluster}< \xi_\mathrm{int}$.    In contrast, 
the coherence length in the neutron superfluid outside the cluster
is estimated to be $\xi_\mathrm{ext}=4.5$ fm. The wave length of the
oscillation there is much larger than $\xi_\mathrm{ext}$, thus the 
existence of hydrodynamic AB phonon mode is allowed there.
This argument may be helpful to understand the feature that the
AB phonon mode does not penetrate inside the cluster.

\subsection{Systematics of dipole AB phonon}

\begin{figure}[tbh]
\begin{center}
\includegraphics[width=0.470\textwidth,keepaspectratio]{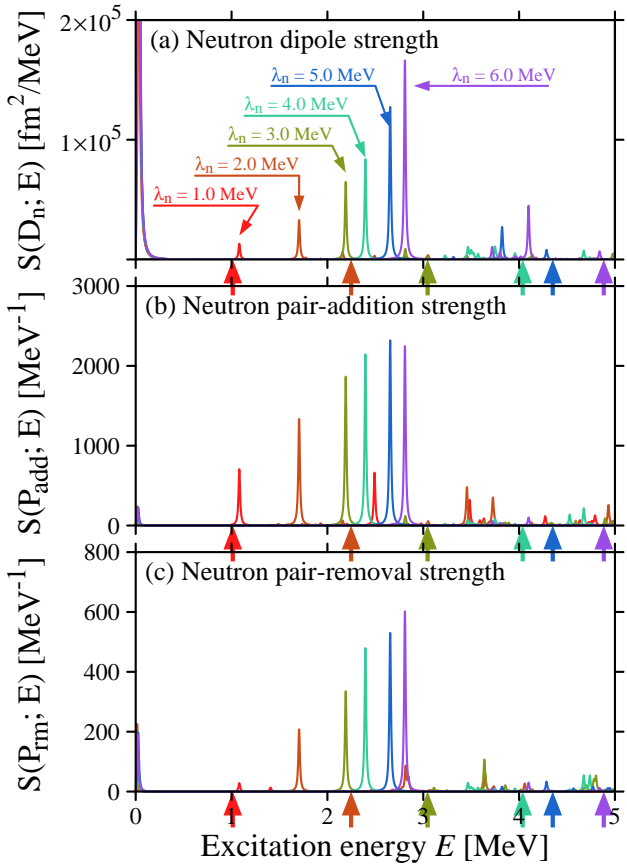}
\caption{(Color online) 
Strength functions  (a)  $S(D_n;E)$, (b) $S(P_\mathrm{add};E)$, 
and (c)  $S(P_\mathrm{rm};E)$ for $Z=28$ system with $\lambda_n =$ 1.0 MeV (red), 
2.0 MeV (brown), 3.0 MeV (yellow green), 4.0 MeV (blue green), 5.0 MeV (blue), and 6.0 MeV (purple).
Colored arrows indicate the threshold energy $2\Delta_\mathrm{ext}$ for corresponding $\lambda_n$.
See text for details.
}
\label{ABmode.dipole.lambda}
\end{center}
\end{figure}

We shall discuss systematic behaviour of the dipole AB phonon mode  by varying $\lambda_n$ and $Z$.
We also explore a possibility of higher harmonics of the AB phonon.  

\begin{figure}[!tbh]
\begin{center}
\includegraphics[width=0.4700\textwidth,keepaspectratio]{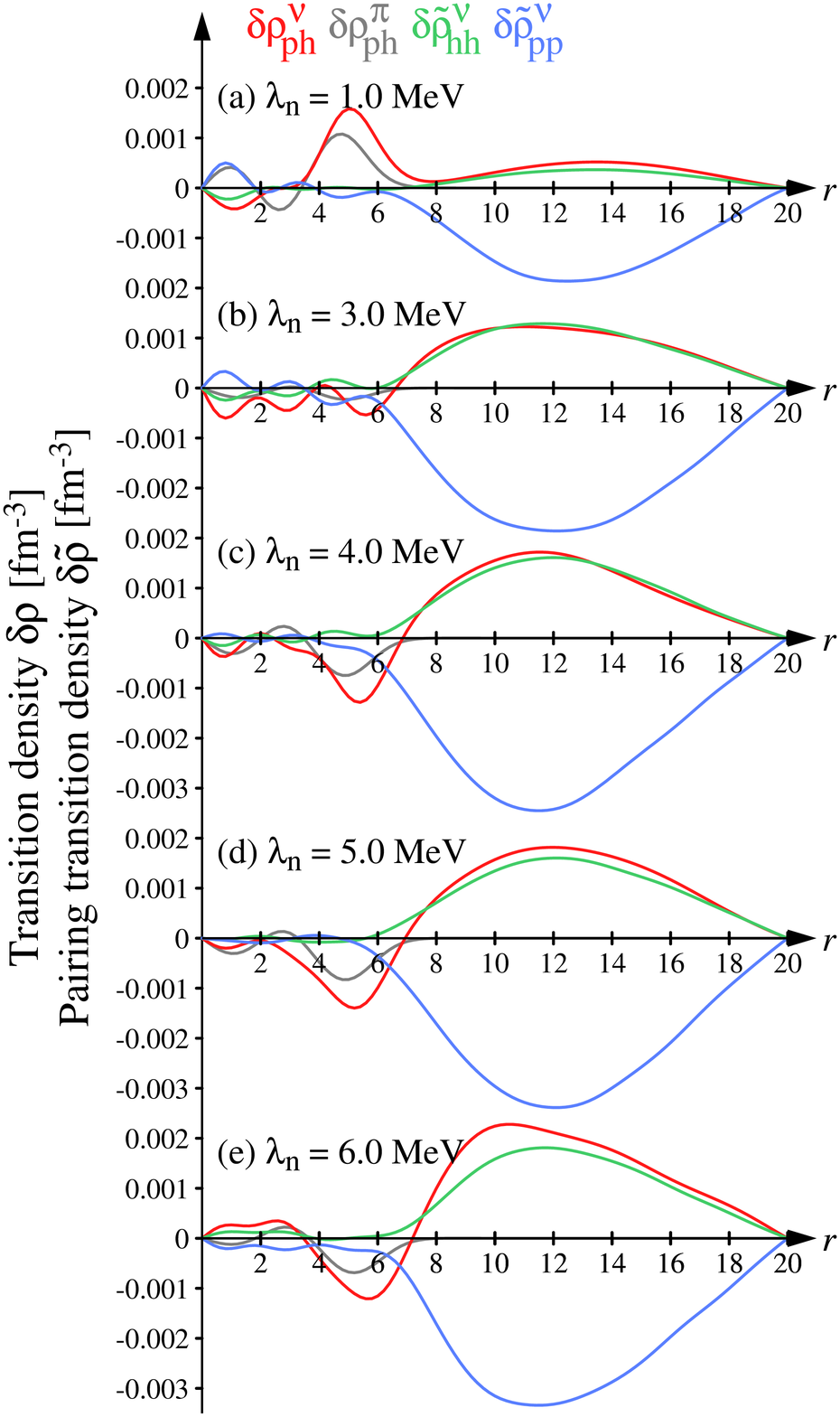}
\caption{(Color online) 
Transition densities 
$\delta\rho^\nu_\mathrm{ph}$, $\delta\rho^\pi_\mathrm{ph}$, 
$\delta\tilde{\rho}^\nu_\mathrm{pp}$, and $\delta\tilde{\rho}^\nu_\mathrm{hh}$ of the dipole
AB phonon mode for $Z=28$ system with 
 $\lambda_n=$ (a) 1.0 MeV, (b) 3.0 MeV, (c) 4.0 MeV, (d) 5.0 MeV and (e) 6.0 MeV.
See text for details.
}
\label{drho.lambda}
\end{center}
\end{figure}

\begin{table*}
\caption{Neutron chemical potential $\lambda_n$, neutron density  $\rho_\mathrm{ext}$ and neutron pair gap $\Delta_\mathrm{ext}$
outside the cluster, 
and excitation energy $E_\mathrm{AB}$ [MeV], neutron dipole strength $B(D_n)$, proton dipole strength $B(D_p)$, 
neutron pair-addition strength $B(P_\mathrm{add})$, and neutron pair-removal strength $B(P_\mathrm{rm})$ 
of the dipole AB phonon mode obtained for  $Z=28$ system with $\lambda_n = 1.0 - 6.0$ MeV.}
\label{table1}
\begin{ruledtabular}
\begin{tabular}{cccccccc}
$\lambda_n$ [MeV] & $\rho_\mathrm{ext}$ [fm$^{-3}$]  & $\Delta_\mathrm{ext}$ [MeV]  & $E_\mathrm{AB}$ [MeV] & $B(D_n)$ [fm$^2$] & $B(D_p)$ [fm$^2$] & $B(P_\mathrm{add})$ & $B(P_\mathrm{rm})$ \\
\hline
  1.0  &  4.67 $\times$ 10$^{-4}$   &   0.51  &  1.08   &  1.17 $\times$ 10$^{3}$ &  -    &  62.3 &  2.49  \\
  2.0  &  1.67 $\times$ 10$^{-3}$   &   1.12  &  1.71   &  2.91 $\times$ 10$^{3}$ &  -    & 117.8 & 18.33  \\
  3.0  &  3.25 $\times$ 10$^{-3}$   &   1.52  &  2.19   &  5.72 $\times$ 10$^{3}$ & 0.144 & 164.7 & 29.59  \\
  4.0  &  5.58 $\times$ 10$^{-3}$   &   2.02  &  2.40   &  7.41 $\times$ 10$^{3}$ & 0.464 & 189.8 & 42.40  \\
  5.0  &  8.23 $\times$ 10$^{-3}$   &   2.18  &  2.66   &  1.13 $\times$ 10$^{4}$ & 0.611 & 204.8 & 46.76  \\
  6.0  &  1.17 $\times$ 10$^{-2}$   &   2.44  &  2.81   &  1.47 $\times$ 10$^{4}$ & 0.527 & 198.5 & 53.13  \\
\end{tabular}
\end{ruledtabular}
\end{table*}

We first look into dependence on the neutron chemical potential $\lambda_n$. 
In Fig.~\ref{ABmode.dipole.lambda} we plot the strength functions 
$S(D_n;E)$, $S(P_\mathrm{add};E)$, and $S(P_\mathrm{rm};E)$ for 
$Z=28$ systems with $\lambda_n= 1.0, 2.0, \cdots,  6.0$ MeV. 
It is seen in all the cases that low-energy collective states having large strengths 
emerge around 1 -- 3 MeV of excitation energy.  The transition
densities of these collective states are shown 
in Fig.~\ref{drho.lambda} for $\lambda_n=1.0,3.0, \cdots, 6.0$ MeV, 
and it is found that they all share  the basic characteristics of the AB phonon mode 
found in the case $\lambda_n=2.0$ MeV. 

\begin{figure}[tbp]
\begin{center}
\includegraphics[width=0.470\textwidth,keepaspectratio]{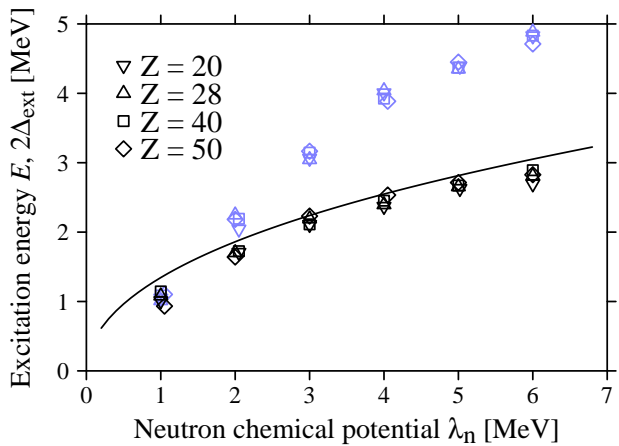}
\caption{(Color online) Excitation energy of the dipole AB phonon mode for 
$Z=$ 20 ($\bigtriangledown$), 28 ($\bigtriangleup$), 40 ($\square$), and 50 ($\diamond$) 
plotted as a function of neutron chemical potential $\lambda_n$. Blue symbols are
the threshold energy
$2\Delta_\mathrm{ext}$.
Hydrodynamic estimate of the excitation energy is also plotted with solid line. See text for details.
}
\label{ABmode.Zall}
\end{center}
\end{figure}

Concerning the $\lambda_n$-dependence,  the excitation energy increases with increase of 
$\lambda_n$. It is seen also that
the peak height of the neutron strength functions $S(D_n;E)$, $S(P_\mathrm{add};E)$ and $S(P_\mathrm{rm};E)$,
and hence the transition strengths develop with increasing $\lambda_n$.
The excitation energy and the transition strengths are listed in Table \ref{table1}, and 
 the $\lambda_n$-dependence of the excitation energy is shown
in Fig.~\ref{ABmode.Zall}. 

The $\lambda_n$-dependence of the excitation energy can be assessed
in a simple way. For this purpose
we consider a simple hydrodynamic description of AB phonon, and suppose that
it is modified by the presence of the cluster. Namely we assume a standing
wave taking place outside the cluster. Then
the phonon amplitude 
may be expressed as $\phi(r) \propto j_1(qr) + A n_1(qr)$ in terms of the spherical Bessel and Neumann 
functions together with the boundary condition $\phi(r)=0$
at the box edge $r=R_\mathrm{box}=20$ fm and with the approximate node at $r\approx 7$ fm located in the
surface region of cluster. This determines the wave number $q=0.265$ fm$^{-1}$ for this standing wave.
Combination of  the phonon dispersion relation $\omega = c_\mathrm{ph} q$ and 
the hydrodynamic estimate of the phonon 
velocity $c_\mathrm{ph}=\sqrt{(\partial\lambda_n/\partial\rho)\rho/m}$
gives an estimated excitation energy $\hbar\omega_\mathrm{hyd}$
\footnote{In this estimate, we use $\rho(\lambda_n)$ evaluated for 
uniform neutron matter  with the same Skyrme functional SLy4. Note also
that according to Ref. ~\cite{Martin14} the
phonon dispersion $\omega = c_\mathrm{ph} q$ with the hydrodynamic phonon velocity $c_\mathrm{ph}$
may be realized for $\hbar\omega$ smaller than $2\Delta$, but deviate as $\omega < c_\mathrm{ph} q$
for $\hbar\omega \gesim \Delta$. Therefore
$\hbar\omega_\mathrm{hyd}$ gives a rough upper estimate of the excitation energy.},
which is
plotted as solid curve in Fig.~\ref{ABmode.Zall}.
 We find reasonably good agreement between
the microscopically evaluated excitation energy and this simple estimate.  If we suppose 
a plane wave phonon $\phi(r) \propto j_1(qr)$ in uniform neutron superfluid, 
i.e. without influence of the cluster, then $q=0.224$ fm$^{-1}$
and the comparison becomes worth. All these indicate that the collective state
under discussion is described  in terms of the modified hydrodynamic picture
in a zero-th order approximation. 
An exceptional example is the case of  $\lambda_n=1.0$ MeV.
The collectivity may not be developed well in this case
as seen in the transition densities (Fig.~\ref{drho.lambda}(a)) where the simple phonon
behaviour is partially broken, and in the excitation energy which is slightly larger than 
$2\Delta_\mathrm{ext}$.  
Note however that we would have a well developed AB phonon mode even in the case of $\lambda_n=1$ MeV 
if we took such a large Wigner-Seitz cell that gives small $q$ and small excitation energy
$ \ll 2\Delta_\mathrm{ext}$.

In addition to the zero-th order picture discussed above,  an interesting behaviour of the
transition densities is seen in the cases of $\lambda_n \gesim 4.0$ MeV 
(Figs.~\ref{drho.lambda}(c)-(e))  in the region 
$r=3-8$ fm around the surface of the cluster. In these cases,  
the neutron particle-hole transition density  $\delta\rho^\nu_\mathrm{ph}(r)$
exhibits a clear node at $r \approx 7$ fm, and it has sizable amplitude
in the region   $r \approx 3-7$ fm with opposite sign to the amplitude outside $r\gesim 7$ fm.
It is seen also that the proton particle-hole amplitude $\delta\rho^\pi_\mathrm{ph}(r)$ has
finite amplitude in this region with coherence to $\delta\rho^\nu_\mathrm{ph}(r)$,
exhibiting some resemblance to the displacement motion of the cluster (see Fig.~\ref{CM.ABmode.LED}(b)).
It is as if oscillation of superfluid neutrons outside the surface repels the 
cluster, and a core part $r\lesim 7$ fm of the cluster
receives a recoil motion in the opposite direction.
However, the
amplitude of the `recoil motion'  is small as deduced from
the transition strength for the proton dipole operator $\sim O(10^{-1})$ fm$^{2}$ (
Table~\ref{table1}), which is much smaller than the single-particle values $\sim R^2_\mathrm{nucl} 
\sim O(10^{1})$ fm$^{2}$. 
It is interesting to note that this characteristic behaviour of the transition
densities is similar to those of 
the pygmy dipole resonance and the low-energy dipole
excitations~\cite{Paar07} of isolated neutron-rich nuclei. 
In the latter case, however, the oscillation in the outer area is not that of
neutron superfluid, but neutron skin or extended tail of neutron
density associated with weakly bound neutrons.

\begin{figure}[tb]
\begin{center}
\includegraphics[width=0.470\textwidth,keepaspectratio]{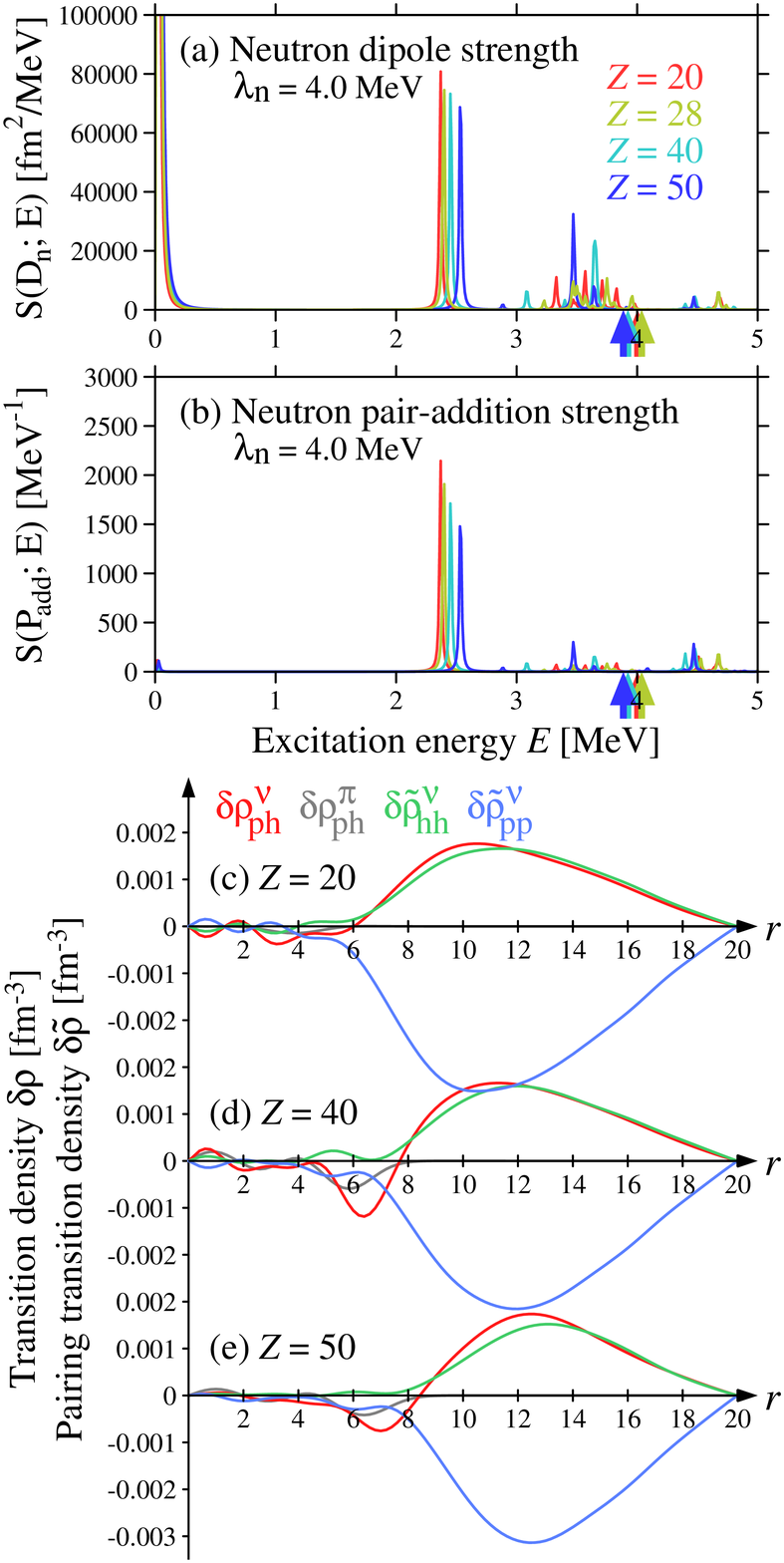}
\caption{(Color online) 
Uppers: Strength functions  (a)  $S(D_n;E)$ and 
(b)  $S(P_\mathrm{add};E)$ for $Z=$ 20 (red), 28 (light green), 40 (cyan), and 50 (blue) systems with fixed $\lambda_n=$ 4.0 MeV.
Colored arrows indicate the threshold energy $2\Delta_\mathrm{ext}$ for corresponding $Z$.
Lowers: Transition densities 
$\delta\rho^\nu_\mathrm{ph}$, $\delta\rho^\pi_\mathrm{ph}$, 
$\delta\tilde{\rho}^\nu_\mathrm{pp}$, and $\delta\tilde{\rho}^\nu_\mathrm{hh}$ 
of the dipole AB phonon for  
(c) $Z=20$, (d) $Z=40$, and (e) $Z=50$ systems. See text for details.
}
\label{Zdep}
\end{center}
\end{figure}

We shall briefly discuss dependence on the proton number $Z$ of the cluster.
We have performed  calculations for systems with different proton numbers $Z=20,28,40,50$,
corresponding to Ca, Ni, Zr, and Sn, respectively, and with various $\lambda_n$.
We found in all the cases the AB phonon mode having the same character as discussed above.
Examples with $\lambda_n=4.0$ MeV and  $Z=20,28,40,50$ are shown
in Figs.~\ref{Zdep}(a)(b), where are plotted the strength functions
 $S(D_n;E)$ and $S(P_\mathrm{add};E)$. The transition densities are shown 
in Figs.~\ref{Zdep}(c)-(e).
It is seen in Figs.~\ref{Zdep}(a)(b) that the excitation energy varies only weakly with the change of the proton number. The transition strengths (the peak height)
are also insensitive to $Z$. This is a reasonable consequence of the fact that the present AB phonon mode is essentially the excitation of neutron superfluid taking place only outside the cluster. 
We note also that the characteristic coupling between the AB phonon mode and the 
recoil motion of the core part of cluster, demonstrated in Figs.~\ref{drho.lambda}(c)-(e), is seen systematically 
for larger values of the neutron chemical potential $\lambda_n \gesim 4-5$ MeV.

\begin{figure}[tbp]
\begin{center}
\includegraphics[width=0.470\textwidth,keepaspectratio]{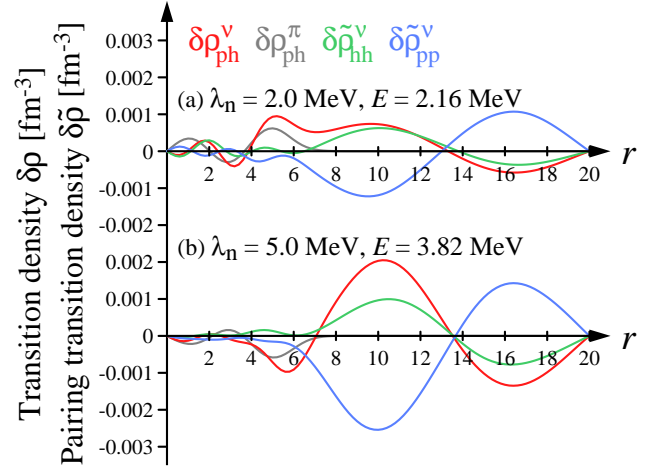}
\caption{(Color online) 
Transition densities
$\delta\rho^\nu_\mathrm{ph}$, $\delta\rho^\pi_\mathrm{ph}$, 
$\delta\tilde{\rho}^\nu_\mathrm{pp}$, and $\delta\tilde{\rho}^\nu_\mathrm{hh}$ 
of the second AB phonon state: 
(a) 2.16 MeV state for $Z=28$ with $\lambda_n =$ 2.0 MeV and 
(b) 3.82 MeV state for $Z=28$ with $\lambda_n =$ 5.0 MeV. 
}
\label{ABmode.higherHarmonics}
\end{center}
\end{figure}

Finally, we explore possibility of higher harmonics of the AB phonon mode, which may 
emerge if the collectivity of the phonon mode is sufficiently strong. In the case of
$\lambda_n=2.0$ MeV and $Z=28$, we find that the second excited state, a very small
peak at $E=2.16$ MeV in Fig.~\ref{neutron.proton.Padd.Premove}, has partly the character of the 
second harmonics of the AB phonon mode.
The transition densities of this state is shown
in Fig.~\ref{ABmode.higherHarmonics}(a), which has
 an oscillation node at $r_\mathrm{node}\approx 13$ fm. (The small transition strengths
are due to cancellation between the amplitudes in two regions $r>r_\mathrm{node}$
and $r<r_\mathrm{node}$.) However, the oscillatory behaviour is 
not as clear as the first AB phonon mode at $E=1.71$ MeV. We deduce that the
collectivity does not develop well as the excitation energy is comparable
to $2\Delta_\mathrm{ext}$. Also for $\lambda_n \ge 3.0$ MeV,
we find a  state interpreted as the second harmonics of the AB phonon mode.
  The excitation energies are $E=$ 2.81, 3.47, 3.82, 4.10 MeV
for $\lambda_n =$ 3.0, 4.0, 5.0, 6.0 MeV. One example is shown in Fig.~\ref{ABmode.higherHarmonics}(b). 
We see a clearer oscillatory profile in these cases, and it may be due to more developed
collectivity,  reflecting the
excitation energy smaller than $2\Delta_\mathrm{ext}$.

\section{Conclusion}

We have investigated the Anderson-Bogoliubov (AB) phonon  (called also superfluid phonon) appearing in the inner crust of neutron stars by employing the density functional description of nucleon many-body collective dynamics.
We consider configurations where spherical clusters are immersed in neutron superfluid, and
formulate the HFB plus QRPA model in a spherical Wigner-Seitz cell.  We adopt the Skyrme energy density functional SLy4 and  the density-dependent delta interaction, which are designed to describe
not only finite nuclei but also the EOS and the pairing properties of neutron matter. 
In the present paper, we have focused on the AB phonon mode with the dipole multipolarity,
relevant to the coupling to the lattice phonon or the displacement motion of clusters, and studied 
how the presence of clusters  influences the AB phonon.
Numerical analysis is performed by varying the neutron chemical potential 
$\lambda_n =1 - 6$ MeV (changing the density of superfluid neutrons)  and the proton
numbers of clusters.

Our model demonstrates systematic emergence of very collective low-energy excitations which
display clearly  typical characteristics of the  AB phonon  in the external region of clusters, i.e.
in surrounding neutron superfluid. However, this AB phonon mode is modified strongly by the presence of clusters. 
The phonon amplitude
is significantly reduced inside the surface of cluster as if  
the AB phonon does not penetrate into clusters. 
 This suggests that
the coupling of the AB phonon to the lattice phonon may be weaker than what is
expected from a simple hydrodynamic description assuming uniform neutron superfluid.
Examining in detail results from various neutron chemical potentials $\lambda_n$, we find 
also that  behaviour  of the AB phonon around the clusters depends on $\lambda_n$ or
the density of surrounding neutron superfluid. At higher densities corresponding to
 $\lambda_n \gesim 4-5$ MeV, the phonon amplitudes display a clear node 
 in the surface area of clusters, as if the AB phonon in neutron superfluid gives 
 recoil motion to core part of the clusters in opposite direction to the AB phonon outside.
This behaviour resembles that of the pygmy dipole excitation of neutron-rich nuclei.
This points to a possibility that 
the coupling between the lattice phonon and 
the AB phonon has non-trivial dependence on $\lambda_n$, and may become
even weaker for $\lambda_n \gesim 4-5$ MeV. 
Quantitative analysis of the phonon-phonon coupling and its impact 
on the thermal conductivity of inner crust will be discussed in forthcoming papers.

\section*{Acknowledgments}

This work is financially supported by Grant-in-Aid for Scientific 
Research on Innovative Areas, No. 24105008, by The Ministry of Education, 
Culture, Sports, Science and Technology, Japan. \\


\begin{thebibliography}{99}

\bibitem{Chamel-Haensel2008} 
N.~Chamel, and P.~Haensel, Living Rev. Relativity, {\bf 11}, 10 (2008).

\bibitem{Haensel-book}
P.~ Haensel, A.~Y.~Potekhin, D.~G.~Yakovlev, {\it Neutron Stars 1: Equation of State and
Structure}, Astrophysics and Space Science Library, vol.326 (Springer, New York, 2007).

\bibitem{Pethick-Ravenhall95} 
C.~J.~ Pethick, D. G. Ravenhall, Annu. Rev. Nucl. Part. Sci. {\bf 45}, 429 (1995).

\bibitem{Anderson-Itoh1975} 
P.~W.~Anderson, N.~Itoh,
Nature, {\bf 256}, (1975).

\bibitem{Alpar1977} 
M.~Ali~Alpar,
Astrophys. J. {\bf 213}, 527 (1977)

\bibitem{Pines-Alpar92} 
D.~Pines and M.~Ali Alpar, in {\it The Structure and Evolution of Neutron Stars},
ed. D. Pines, R. Tamagaki, and S. Tsuruta ( Addison Wesley, 1992), p.7.

\bibitem{Lattimer94}  
J.~M.~Lattimer, K.~A.~van Riper, M.~Prakash, and M.~Prakash,
Astrophys. J. {\bf 425}, 802 (1994).

\bibitem{Gnedin01} 
O.~Y.~Gnedin, D.~G.~Yakovlev, and A.~Y.~Potekhin, 
Mon. Not. R. Astron. Soc. {\bf 324}, 725 (2001).

\bibitem{Shternin07}  
P.~S.~Shternin, D.~G.~Yakovlev, P.~Haensel, and A.~Y.~Potekhin,
Mon. Not. R. Astron. Soc. Lett. {\bf 382}, L43 (2007).

\bibitem{Brown09} 
E.~F.~Brown and A.~Cumming, 
Astrophys. J. {\bf 698}, 1020 (2009).

\bibitem{Pizzochero2002} 
P.~M.~Pizzochero, F.~Barranco, E.~Vigezzi and R.~A.~Broglia,
Astrophys. J. {\bf 569}, 381 (2002).

\bibitem{Monrozeau07} 
C.~Monrozeau, J.~Margueron, and N.~Sandulescu, 
Phys. Rev. C {\bf 75}, 065807 (2007).

\bibitem{Fortin2010} 
M.~Fortin, F.~Grill, J.~Margueron, D.~Page, and N.~Sandulescu,
Phys. Rev. C {\bf 82}, 065804 (2010).

\bibitem{Duncan1998} 
R.~C.~Duncan, Astrophys. J. {\bf 498}, L45 (1998).

\bibitem{Samuelson2007} 
L.~Samuelson, and N.~Andersson,
Mon. Not. R. Astron. Soc., {\bf 374}, 256 (2007).

\bibitem{Andersson09} 
N.~Andersson, K.~Glampedakis, and L.~Samuelsson, 
Mon. Not. R. Astron. Soc. {\bf 396}, 894 (2009).

\bibitem{Aguilera09} 
D.~N.~Aguilera, V.~Cirigliano, J.~A.~Pons, S.~Reddy, and R.~Sharma,
Phys. Rev. Lett. {\bf 102}, 091101 (2009).

\bibitem{Pethick10} 
C.~J.~Pethick, N.~Chamel, and S.~Reddy,
Prog. Theor. Phys. Supple. {\bf 186}, 9 (2010).

\bibitem{Cirigliano11} 
V.~Cirigliano, S.~Reddy, and R.~Sharma,
Phys. Rev. C {\bf 84}, 045809 (2011).

\bibitem{Page-Reddy2012}
D.~Page and S.~Reddy, in {\it Neutron Star Crust},
ed. by C.~Bertulani and J.~Piekarewicz (Nova Science, 2012), p.281

\bibitem{Chamel13} 
N.~Chamel, D.~Page, and S.~Reddy,
Phys. Rev. C {\bf 87}, 035803 (2013);
J. Phys, Conf. Ser., {\bf 665}, 012065 (2016).

\bibitem{Kobyakov13} 
D.~Kobyakov and C.~J.~Pethick, 
Phys. Rev. C {\bf 87}, 055803 (2013)

\bibitem{Kobyakov14} 
D.~Kobyakov and C.~J.~Pethick, 
Phys. Rev. Lett. {\bf 112}, 112504 (2014).

\bibitem{Martin14} 
N.~Martin and M.~Urban,
Phys. Rev. C {\bf 90}, 065805 (2014).

\bibitem{Anderson58}
P.~W.~Anderson, 
Phys. Rev. {\bf 112}, 1900 (1958).

\bibitem{Bogoliubov59}
N.~N.~Bogoliubov, V.~V.~Tolmachev, and D.~V.~Shirkov,
A New Method in the Theory of Superconductivity (Academy of Science, Moscow, 1958, New York, 1959).

\bibitem{Galitskii58}
V.~M.~Galitskii, 
JETP {\bf 34}, 1011 (1958).

\bibitem{Chamel10} 
N.~Chamel, S.~Goriely, J.~M.~Pearson, and M.~Onsi,
Phys. Rev. C {\bf 81}, 045804 (2010) 

\bibitem{Sandulescu04a} 
N.~Sandulescu, N.~V.~Giai, and R.J.~Liotta,
Phys. Rev. C {\bf 69}, 045802 (2004).

\bibitem{Sandulescu04b} 
N.~Sandulescu, 
Phys. Rev. C {\bf 70}, 025801 (2004).

\bibitem{Baldo2005} 
M.~Baldo, U.~Lombardo, E.~E.~Sperstein, and S.~V.~Tolokonnikov,
Nucl. Phys. {\bf A750}, 409 (2005).

\bibitem{Baldo2006} 
M.~Baldo,  E.~E.~Sperstein, and S.~V.~Tolokonnikov,
Nucl. Phys. {\bf A775}, 235 (2006).


\bibitem{Grill11} 
F.~Grill, J.~Margueron, and N.~Sandulescu,
Phys. Rev. C {\bf 84}, 065801 (2011).

\bibitem{Pastore11} 
A.~Pastore, S.~Baroni, and C.~Losa,
Phys. Rev. C {\bf 84}, 065807 (2011)

\bibitem{Pastore12} 
A.~Pastore,
Phys. Rev. C {\bf 86}, 065802 (2012).

\bibitem{Barranco1998} 
F.~Barranco, R.~A.~Broglia, H.~Esbensen, and E.~Vigezzi,
Phys. Rev. C {\bf 58}, 1257.

\bibitem{Khan05} 
E.~Khan, N.~Sandulescu, and N.~V.~Giai, 
Phys. Rev. C {\bf 71}, 042801 (R) (2005).

\bibitem{Grasso08} 
M.~Grasso, E.~ Khan, J.~Margueron, N.~V.~Giai,
Nucl. Phys. {\bf A 807}, 1 (2008).

\bibitem{Gori2004} 
G.~Gori, F.~Ramponi, F.~Barranco, R.~A.~Broglia, G.~L.~Colo, D.~Sarchi,
Nucl. Phys. {\bf A731}, 401 (2004).

\bibitem{Baroni2010} 
S.~Baroni, A.~Pastore, F.~Raimondi, F.~Barranco, R.~A.~Broglia, and E.~Vigezzi,
Phys. Rev. C {\bf 82}, 015807 (2010).

\bibitem{Nakatsukasa2016}
T.~Nakatsukasa, K.~Matsuyanagi, M.~Matsuo, and K.~Yabana,
Rev. Mod. Phys. {\bf 88}, 045004 (2016).

\bibitem{Matsuo01} 
M.~Matsuo, 
Nucl. Phys. {\bf A 696}, 371 (2001); 
Prog. Theor. Phys. Suppl. {\bf 146}, 110 (2002).

\bibitem{Matsuo05} 
M.~Matsuo, K.~Mizuyama, and Y.~Serizawa, 
Phys. Rev. C {\bf 71}, 064326 (2005).

\bibitem{Serizawa09} 
Y.~Serizawa and M.~Matsuo, 
Prog. Theor. Phys. {\bf 121}, 97 (2009).

\bibitem{Matsuo10} 
M.~Matsuo and Y.~Serizawa,
Phys. Rev. C {\bf 82}, 024318 (2010).

\bibitem{Negele73}
J.~W.~Negele and D.~Vautherin, 
Nucl. Phys. {\bf A 207}, 298 (1973).

\bibitem{SLy4}
E.~Chabanat, P.~Bonche, P.~Heenen, J.~Meyer, and R.~Schaeffer, 
Nucl. Phys. {\bf A 635}, 231 (1998). 

\bibitem{APR}
A.~Akmal, V.~R.~Pandharipande, and D.~G.~Ravenhall, 
Phys. Rev. C {\bf 58}, 1804 (1998).

\bibitem{Matsuo07}
M.~Matsuo, Y.~Serizawa, and K.~Mizuyama, 
Nucl. Phys. {\bf A 788}, 307c (2007). 

\bibitem{Matsuo06} 
M.~Matsuo, 
Phys. Rev. C {\bf 73}, 044309 (2006).

\bibitem{Khan04} 
E.~Khan, N.~Sandulescu, N.~V.~Giai, and M.~Grasso,
Phys. Rev. C {\bf 69}, 014314 (2004).

\bibitem{Khan09} 
E.~Khan, M.~Grasso, and J.~Margueron, 
Phys. Rev. C {\bf 80}, 044328 (2009).

\bibitem{Paar07} 
N.~Paar, D.~Vretenar, E.~Khan, and G.~Col\`{o},
Rep. Prog. Phys. {\bf 70}, 691 (2007).

\bibitem{Papakonstantinou13}
P.~Papakonstantinou, J.~Margueron, F.~Gulminelli, and Ad.~R.~Raduta, 
Phys. Rev. C {\bf 88}, 045805 (2013).

\bibitem{Pethick-Smith2002}
C.~J.~Pethick and H.~Smith,
{\it Bose-Einstein Condensation in Dilute Gases},
 (Cambridge University Press, Cambridge, 2002).

\end{thebibliography}
\end{document}